\RequirePackage{fix-cm}
\documentclass[smallextended]{svjour3}       
\smartqed  
\usepackage{graphicx}
\usepackage{multirow}
\usepackage{color}

\begin{document}

\title{Effect of different patient peak arrivals on an Emergency Department via discrete event simulation
}

\titlerunning{Effect of different patient peak arrivals on an ED via Discrete Event Simulation}        

\author{Giordano Fava
        \and
        Tommaso Giovannelli~\href{https://orcid.org/0000-0002-1436-5348}{\includegraphics[scale=0.5]{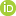}}
        \and
        Mauro Messedaglia
        \and
        Massimo Roma~\href{https://orcid.org/0000-0002-9858-3616}{\includegraphics[scale=0.5]{orcid.png}}
}

\authorrunning{G. Fava, T. Giovannelli, M. Messedaglia, M. Roma} 

\institute{Giordano Fava, Tommaso Giovannelli, Massimo Roma \at
                Dipartimento di Ingegneria Informatica Automatica e Gestionale ``A. Ruberti'', SAPIENZA Universit\`a di Roma, via Ariosto, 25 -- 00185 Roma, Italy.
              \email{giordanotu.fava@gmail.com, giovannelli@diag.uniroma1.it, roma@diag.uniroma1.it}           
           \and
           Mauro Messedaglia \at
              ACTOR Start up of SAPIENZA Universit\`a di Roma,
              via Nizza 45, 00198 Roma, Italy.
              \email{mauro.messedaglia@gmail.com}
}

\date{}

\maketitle

\begin{abstract}
Emergency Departments (EDs) overcrowding is a well recognized worldwide phenomenon. The consequences range from long waiting times for visit and treatment of patients, up to life-threatening health conditions.
The international community is devoting greater and greater efforts to analyze this phenomenon aiming at reducing waiting times, improving the quality of the service.
Within this framework, we propose a Discrete Event Simulation (DES) model to study the patient flows through a medium--size ED located in a region of Central Italy recently hit by a severe earthquake. In particular, our aim is to simulate unusual ED conditions, corresponding to critical events (like a natural disaster) which cause a sudden spike in the number of patient arrivals. The availability of detailed data concerning the ED processes enabled to build an accurate DES model and to perform extensive scenario analyses. The model provides a valid decision support system for the ED managers also in defining specific emergency plans to be activated in case of 
mass casualty disasters.

\keywords{Emergency Department \and Discrete Event Simulation \and Patient flow \and Overcrowding \and Patient peak arrivals}
\end{abstract}

\section{Introduction}
Emergency Department (ED) belongs to the Emergency Medical Services (EMS) which represent the most important healthcare services, considering that they affect people's lives (see the recent survey paper \cite{Aringhieri.2017} for a complete review on EMS). In particular,
the task of an ED is providing healthcare services for people who need urgent medical treatments. An ED is open 365 days a year and 24 hours a day and people with different urgency arrive requiring treatment. Since the services delivered by an ED are time--critical, the main issue concerns the response time. Unfortunately, the well known and increasing problem of the overcrowding tends to enlarge the waiting times, endangering the life of critical patients. Today the overcrowding is an international problem widely considered in the specific literature (see e.g. \cite{hoot.aronsky:08,Weiss.2004,Weiss.2006} and the references reported therein). In particular, according to \cite{hoot.aronsky:08}, possible causes of overcrowding are insufficient staff, shortcomings of the structures, flu season, request of nonurgent treatments, unavailability of hospital beds. Besides treatment delays, other possible consequences of overcrowding are reduced quality of the services and higher patient mortality, ambulance diversion, growing number of patients which leave without being visited, greater expenses for the service provider due to longer patient stay. Of course, these critical issues are greatly amplified whenever  mass casualty disasters occur. In this case, the rate of patient arrivals suddenly increases and some different tactical and strategic decisions must be adopted to ensure timely treatments.
\par
In this paper we propose the use of the DES based approach for analyzing the operation of an ED of a medium--size hospital located in an Italian region where the recent earthquakes have put a strain on the EDs of the area. For such an ED it is crucial to assess the impact of unusual/critical events which cause peak arrivals in order to possibly adopt suited contingency plans. Therefore, we study the effects of spikes in the number of patient arrivals on the ED operation. The availability of detailed data concerning the ED processes allowed us to build an accurate DES model, well reproducing the actual operating modes of the ED. After a complete input analysis, the DES model has been implemented by using {\sf ARENA~15.1} Simulation Software \cite{ARENA,simulationwitharena} which is one of the most commonly used general purpose DES package. Based on flow\-chart modules, it enables to build the simulation model and to perform input analysis, simulation runs and output analysis.
Then the model has been accurately validated in order to guarantee its reliability. Several scenario analyses have been performed, aiming at evaluating the impact of possible changes in the ED operating conditions. We focused on assessing the main Key Performance Indicators (KPIs) of the ED for different scenarios in critical conditions. In particular, we simulated unusual ED conditions, corresponding to spikes in the number of patient arrivals. These patient surges can be of different patterns and they seriously affect the ED operation. We experienced some artificial scenarios we have specially created trying to reproduce really occurred conditions. The model we propose is very helpful for the ED managers both in the daily management of the resources and also in defining suited emergency plans to be adopted in case of critical emergencies.
\par
The paper is organized as follows: Section~\ref{background} reports some background material and a literature review. In Section~\ref{sec:gen} generalities on operating conditions of Italian EDs are briefly reported. Section~\ref{sec:casestudy} describes the case study of the ED we consider. In Section~\ref{sec:desm} we detail the DES model we propose, along with the input analysis and the model validation.
Section~\ref{sec:results} reports results for the ``as--is'' status along with extensive scenario analyses, mainly focused on peak patient arrivals. Finally, some concluding remarks are included in Section~\ref{sec:conc}.

\section{Background}\label{background}
In the recent years, techniques from Operations Research have been frequently applied for tackling the problem of the overcrowing of EDs. Since 2005 the \emph{National Academy of Engineering} and the \emph{Institute of Me\-di\-cine} in \cite{NAS.2005} highlighted the importance of using tools from Operations Research and Systems Engineering (statistical process controls, queuing theory, mathematical modelling and simulation) in healthcare delivery, in order to improve performance of care processes or units. Among them, \emph{simulation} is considered of fundamental importance for analyzing several healthcare settings (see, e.g. \cite{NAS.2005} and the several examples reported therein and \cite{Almagooshi.2015}). In particular, simulation models have been largely applied to emergency medical service operations (see \cite{Aboueljinane.2013} for a survey).
Several papers in the simulation literature are devoted to study the patient flow through an ED by means of \emph{Discrete Event Simulation} (DES) models (see e.g.
\cite{Aboueljinane.2013,Gul.2015,Joshi.2016,Kuo.2016,Rado.2014,Wong.2016,Zeinali.2015})
and \emph{Agent Based Simulation} (ABS) models (see e.g.
\cite{Aringhieri.2018,Kaushal.2015,Liu.2015,Taboada.2011,Wang.2009}). In particular,
DES has been largely used for studying causes and effects of EDs overcrowding. We refer to the paper \cite{Paul.2010} (and to the many references reported therein) and to the more recent paper \cite{Nahhas.2017}, for a review of simulation studies available in literature devoted to the ED overcrowding phenomenon and for a discussion on the effectiveness of the approaches based on DES models used for tackling this problem.
Among the many papers which deal with ED overcrowding we mention \cite{Wong.2016} where an ED in Hong Kong is simulated in order to assess how modifying the path of the patient clinical process and the level of physician resources affect the performance; \cite{Joshi.2016} where the simulation model is used to understand the process which allows to shorten the waiting times and the length of stay by varying the workload among staff members and by giving nonurgent patients the possibility to return afterwards; the many papers addressing the adoption of fast--track systems in the ED, such as \cite{Kuo.2018} and \cite{Aroua.2017}, which propose to send less urgent patients to specific queues so that they receive the service early, and hence they will be discharged earlier and the ED will become less crowded; \cite{Whitt.2017} where a in-depth study on patients interarrivals and lenght of stay in and ED located in Israel is reported; \cite{Daldoul.2018} where a stochastic model is considered in order to reduce the average total patient waiting time in an university hospital.
From the wide literature on the topic, it clearly emerges that a study based on DES provides important insights into ED overcrowding. Moreover, as well known, a simulation based study can be also combined with optimization tools in order to determine which setting results the best, once one or more objectives (to be minimized or maximized) are defined. The resulting \emph{Simulation Optimization} approach has been also applied in healthcare contexts (see e.g. \cite{Chanchaichujit:2019,Granja:2014,optl2016,ieee-tase2016,Zhang:2019}) and, in particular, in dealing with ED
(see e.g. \cite{Ahmed.2009,Diefenbach.2011,Guo.2017,Guo.2016}). For instance, in \cite{Guo.2017} a simulation model for the ED of a public hospital in Hong Kong
is built and integrated with an optimization tool in order to find an optimal medical staff configuration to minimize the total labor cost given
the service quality requirement; in \cite{Diefenbach.2011} the patient flow through an Australian ED is studied and optimized on the basis of the bed configuration.
\par
Moreover, interesting papers are devoted to prevent and predict strain situations in EDs (see e.g. \cite{Kadri:2014}) and
to study the effect of spikes in the patient arrivals due to disaster or extreme events. A comprehensive review highlighting the importance of simulation to improve ED policies in case of critical conditions 
can be found in \cite{Gul.2015}. Furthermore, in \cite{Gul.2015b} a DES model is proposed to study disaster scenarios corresponding to a patient flow surge for an ED located in an earthquake area in Istanbul. The aim is to enable early preparedness of ED resources to overcome bottlenecks due to critical situations. Finally, we mention \cite{Xiao.2012} where the patient workflow through an ED located in Western New York during extreme conditions is studied. A framework to reconfigure the workflow is proposed aiming at improving the overall management of the patient flow.

\section{Generalities on EDs and Italian guidelines}\label{sec:gen}
The ED consists of two categories of stakeholders: \emph{physicians} and \emph{nurses}, that are the human resources having different responsibilities, and \emph{patients}, who need a specific care. Moreover, there are physical resources, such as beds, machineries, stretchers and so forth, necessary to host and visit patients during their stay. Each patient arrived to the ED goes through different clinical paths. This flow comprises several steps which generally consist in the triage, whose aim is to assign an urgency code to every patient, medical visits and examinations and, finally, the leaving of the ED, which may involve diverse kinds of discharge.
\par
As first step, a triage tag is assigned to every incoming patient, in order to determine the priority of treatment. Different systems of classification are usually adopted. The most commonly used scale in Italy is reported in Table~\ref{tab:colourcodes}.
\begin{table}[htbp]
\caption{Colour coding scheme for triage of incoming patients.} \label{tab:colourcodes}
\begin{center}
\begin{tabular}{|l|l|}
  \hline
\multirow{2}{*}{\sc Red tag} & Very critical, danger of life. \\
 & The patient must be visited immediately \\ \hline \hline
\multirow{2}{*}{\sc Yellow tag} & Fairly critical, high risk. The patient \\
 & should be visited as soon as possible \\ \hline \hline
\multirow{2}{*}{\sc Green tag} & Minor injury, no risk of conditions \\
 & worsening. The treatment can be delayed. \\ \hline \hline
\multirow{2}{*}{\sc White tag} & No injury, minimal pain with no risk \\
& features. The treatment can be deferred. \\ \hline
\end{tabular}
\end{center}
\end{table}
Moreover, in some regions a blue tag is also used as intermediate case between green and yellow tags. In some countries more fine-grained (sometimes numerical valued) scales are adopted. In order to guarantee more appropriate clinical paths and following the main current international scientific evidence, the Italian Ministry of Health is going to adopt new guidelines. They are based on a new triage which considers five numerical urgency codes (see \cite{triage.2001,accordo.2013}) as detailed in Table~\ref{tab:numericcodes}.
\begin{table}[htbp]
\caption{Numeric coding scheme for the triage of incoming patients.} \label{tab:numericcodes}
\begin{center}
\begin{tabular}{|l|l|}
  \hline
  \sc{Code~1} & Very critical conditions, immediate treatment \\ \hline\hline
  \sc{Code~2} & Fairly critical, high level of risk \\ \hline \hline
  \sc{Code~3} & Not very critical, no risk of worsening \\ \hline \hline
  \sc{Code~4} & Not critical, acute but not serious \\ \hline \hline
  \sc{Code~5} & Not critical, not serious, not acute.  \\ \hline
\end{tabular}
\end{center}
\end{table}
This classification closely resembles the Emergency Severity Index (ESI) adopted in US which is based on an algorithm that rapidly yields grouping of patients into five classes, as described in \cite{Gilboy.2012}.
\par
After assigned the triage code, the nurse activates the more appropriate Diagnostic Therapeutic Care Path (DTCP). In particular, a patient can be sent: 1) to an ED room, 2) to outpatient facilities, 3) toward a \emph{``Fast Track''}, 4) to the \emph{``See and Treat''} service. The Fast Track and See and Treat are novel services recently introduced in order to reduce the waiting times, the Length Of Stay (LOS) in the ED, as well as the percentage of patients who Leave Without Being Seen (LWBS).
The patients directed to ED rooms, follow different clinical pathways which include medical examination and diagnostic tests up to the definition of the outcome. The patient flow inside ED rooms is very complex, due to the many and different specific needs (often even difficult to identify in short time) and the high variability of medical conditions of the incoming patients. Moreover, the flow is also strongly affected by the availability of the resources such as staff on duty, number of rooms and machineries dedicated to different services, capacity of holding areas, beds for hospitalization.
\par
The ED process is usually characterized by the following outcomes: \emph{discharged home} with reliance, if necessary, on territorial structures which provide control at outpatient facilities; \emph{hospitalization} at an hospital ward (if a bed is available) or \emph{transfer} to another hospital; admission to the \emph{Short Stay Unit} (SSU) (whenever such unit exists). The SSU is an inpatient unit attached to the ED, managed under the clinical governance of the ED staff, designed for the short term treatment, observation, assessment and reassessment of patients. 
When a patient is discharged at the end of his clinical pathway, a physician must assign an exit code in order to identify the patient's clinical severity level, similarly to that assigned at the triage.
\par
Unfortunately, very frequently, the great number of incoming patients leads to the overcrowding of an ED. It is a world phenomenon and it is well perceived by the stakeholders: long patients waiting times before the medical examination, excessive number of patients in the ED, high percentage of patients who LWBS are clear indications of such problem. In order to give a formal assessment of the degree of overcrowding, some measures have been proposed. They enable to monitor the state of the ED, describing the current situation and they can also work as alarm bells to avoid reaching a critical level.  The most commonly used are:
the Real Time Emergency Analysis of Demand Indicators (READI), the Emergency Department Work Index (EDWIN), the Work Score, the National Emergency Department Overcrowding Scale (NEDOCS). They are continuous valued indicators computed on the basis of some operational variables which enable to quantify the degree of overcrowding of an ED (see \cite{Hoot.2007} and the references reported therein for the definition of these methods of measurement). However, the study reported in \cite{Hoot.2007} showed that none of these measures actually provides a reliable predictive analysis at a low percentage of false warning.
\par
In order to analyze (and to possibly prevent) the overcrowding phenomenon, it is necessary to detect the time spent by the patient inside the ED, during the different phases of the whole process. To this aim, novel Italian guidelines recommend to monitor the times of the clinical pathway in relation to the assigned priority codes.
In light of these guidelines, a great interest is shown by the ED managers in tools which enable to perform scenario analysis, like those provided by DES. The aim is to assess how the main KPIs change after possible redesigning of ED patient flows and changing of the model of care.
\par
A great and increasing interest concerns the simulation modelling for studying the impact of patient arrival surges caused by some disaster. The related Italian guidelines state specific measures to be adopted in order to efficiently tackle such situations. In particular, the so called
``Internal Emergency Plan for Massive Inflow of Injured''\footnote{In Italian: PEIMAF, Piano di Emergenza Interno per il Massiccio
Afflusso di Feriti} has been issued in recent years.
In this plan, different critical levels have been provided, and suited operative measures are indicated to reallocate ED human and physical resources, whenever it is activated due to critical events. Moreover, low complexity patients can be addressed to outpatient facilities to enable the ED staff to timely deliver the most urgent treatments.

\section{The case study: the ED of the ``E. Profili'' Fabriano hospital}\label{sec:casestudy}
In this section we detail our case study concerning the ED of the ``E. Profili'' Fabriano (Ancona) hospital. This hospital is located in the Italian region of Marche and the catchment area covers about 48000 inhabitants. Every year about 27000 patients arrive to the ED requiring medical assistance, hence it can be considered of medium dimension. A detailed understanding of the ED operation was gained through process mapping performed along with the ED staff. In the sequel we report a brief description of the ED rooms and staff; moreover, we summarize the patient flows through the ED.
This ED is composed by
\begin{itemize}
\item[$\bullet$] a \emph{triage area}, where a nurse assigns the colour tag at each incoming patient and it can host one patient at a time;
\item[$\bullet$] a \emph{waiting room} where patients queue for the triage and (after the triage) they wait for the medical examination;
\item[$\bullet$] three areas for the medical treatment: 
\begin{itemize}
\item the \emph{green area}, for green and white tagged patients,
\item the \emph{yellow area}, for yellow tagged patients,
\item the \emph{shock room} for red tagged patients;
\end{itemize}
\item[$\bullet$] an \emph{holding area};
\item[$\bullet$] a \emph{Short Stay Unit (SSU)}.
\end{itemize}
As regards the areas for the medical treatment, the shock room is the most equipped one, and two critical patients can be hosted simultaneously. In the green area and in the yellow area one seat is available. During the night (9.00 p.m. -- 8.00 a.m.) only the shock room and the green area are in operation, so that also yellow tagged patients are visited in the green area.
\par
As regards the ED staff, physicians and nurses are on duty according to the shifts reported in Tables~\ref{tab:medshifts} and in Table~\ref{tab:nurseshifts}. Other staff members can be engaged in particular emergency.
\begin{table}[htbp]
\caption{Number of physicians on duty in the weekdays (WD) and in the public holidays (PH).}\label{tab:medshifts}
\centering
\begin{tabular}{l|c|c}
   & WD & PH \\ \hline
  \multicolumn{1}{l|}{Morning (8.00 a.m. -- 2.00 p.m.)} & 2 & 1 \\ 
  \multicolumn{1}{l|}{Afternoon (2.00 p.m. -- 9.00 p.m.)} & 2 & 1 \\ 
  \multicolumn{1}{l|}{Night (9.00 p.m. -- 8.00 a.m.)} & 1 & 1 \\
  \hline
\end{tabular}
\end{table}
\begin{table}[htbp]
\caption{Number of nurses on duty each day.}\label{tab:nurseshifts}
\centering
\begin{tabular}{l|c}
  \hline
  \multicolumn{1}{l|}{Morning (7.00 a.m. -- 2.00 p.m.)}    & 3  \\ 
  \multicolumn{1}{l|}{Afternoon (2.00 p.m. -- 10.00 p.m.)} & 3  \\ 
  \multicolumn{1}{l|}{Night (10.00 p.m. -- 7.00 a.m.)}     & 2  \\
  \hline
\end{tabular}
\end{table}
As regards the patient flow, arrivals are by ambulance or autonomously. All the incoming patients are registered at the check--in desk and they are admitted to triage area where a nurse collects patient's health information and assigns the colour tag. Critical patients arriving by ambulance are directly transferred to a medical area for immediate treatment, without going through the triage area. After the triage, a patient waits for the call in the waiting room where the estimated waiting time is displayed on a screen. Then the patient is transferred to an appropriate area inside the ED for medical treatment according to the assigned color tag. In severely urgent cases (red tag), the patient is examined in the shock room for possible immediate interventions. In less severe cases, physicians, after performing health assessment, decide the clinical pathway which must be followed by patient. The pathways can be very differentiated on the basis of the acuity of patient's illness. In many cases, the  physician requires additional examinations for the patient (e.g. clinical laboratory tests, X-ray, EKGs). In other cases, patient is transferred to the SSU for a short observation. Moreover, patients with less serious illness, having a single specialist relevance, are assigned to the fast track service, a specific area of the ED provided by a multidisciplinary team, where timely patient treatment and discharge are ensured. In the ED,
usually one physician and one nurse manage the examination and the treatment of a patient, however in case of severe injuries, the whole ED staff can support them.
\par
\begin{figure}[htbp]
  \centering
    \includegraphics[width=8truecm]{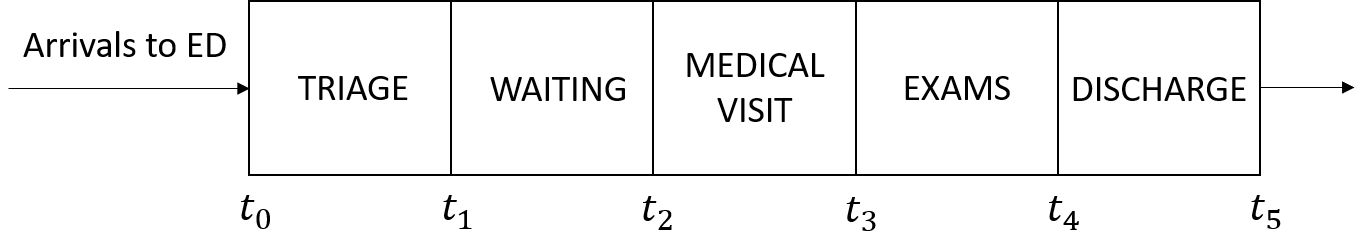}
  \caption{Collected timestamps for the ED process.}\label{fig:times}
\end{figure}

Whenever all the examinations are completed and the related reports have been issued, a reassessment of the patient is performed by the physician of the ED who can require further examinations and/or an additional observation period. At the end of the pathway, the final diagnosis is delivered and patient is discharged from the ED. When a patient is discharged, a physician assigns an exit code in order to identify the patient's clinical severity level, similarly to that assigned at the triage. This implies that the code assigned at the triage on arrival can be confirmed or changed. Furthermore, a detailed description of the outcome must be issued. The outcome is encoded according to the following list:
\begin{description}
\item[]\emph{O1}: patient is discharged home;
\item[]\emph{O2}: patient is discharged home with reliance to outpatient facilities or family physician;
\item[]\emph{O3}: patient is hospitalized at an hospital ward;
\item[]\emph{O4}: patient is transferred to another hospital due to bed unavailability at the appropriate ward of the hospital;
\item[]\emph{O5}: patient refuses hospitalization and leaves the ED despite the medical request;
\item[]\emph{O6}: patient leaves during examinations, i.e. the patient does not complete all the required tests and abandons the ED without informing the staff;
\item[]\emph{O7}: patient leaves without been seen, i.e. the patient abandons the ED waiting room before being examined by a physician (LWBS);
\item[]\emph{O8}: patient dies during the stay at the ED;
\item[]\emph{O9}: patient has arrived deceased to the ED.
\end{description}
In order to perform a complete process mapping of the ED, many interviews have been carried out to the staff (physicians, nurses, managers) and direct observations took place. Moreover, all the available data concerning patient flow during February 2018 have been anonymously collected.
From 00:00 of February 1 to 23:59 of February 28, 2018, the overall number of patients arrived to the ED is 2046. The time--stamps recorded are reported in Figure~\ref{fig:times}. They have been extracted and organized in a suited database. Note that, since the \emph{holding area} and the \emph{SSU} are not subject of our study, these two units of the ED are not specified in our model, and hence not represented in Figure~\ref{fig:times}.
In Table~\ref{tab:triagetag} we report the number and the percentage of color tags assigned at the triage on arrival. Moreover, we also report in Table~\ref{tab:triagetag} the number and the percentage of color tags assigned on discharge (as already noticed, in some cases, the discharge tag, at the end of the clinical pathway can be changed with respect to that assigned at the triage). In the same table the percentage of patient leaving without being seen, for each triage tag, is reported. Finally, the number and the percentage of deceased patients is included in the table, too.
\begin{table}[htbp]
\caption{Number and percentage of color tags assigned at the triage at incoming patients (columns 2-3). Number and percentage of discharge color tags assigned at the discharge (columns 4-5). Percentage of patients LWBS (column 6).}\label{tab:triagetag}
\centering
\begin{tabular}{l|c|c||c|c||c}
& \multicolumn{2}{c||}{TRIAGE TAG} & \multicolumn{2}{|c||}{DISCH. TAG} & \multicolumn{1}{c}{LWBS}\\
\hline
\multicolumn{1}{l|}{\sc White} & 149 & 7.28\% & 171 & 8.36\% &  4.65\%    \\ 
\multicolumn{1}{l|}{\sc Green} & 1448 & 70.77\% & 1612 & 78.78\%& 1.61\%   \\ 
\multicolumn{1}{l|}{\sc Yellow} & 434 & 21.21\% & 243 & 11.88\% & 0.41\%  \\ 
\multicolumn{1}{l|}{\sc Red} & 15 & 0.74\% & 18 & 0.88\%&  - \\ 
\multicolumn{1}{l|}{Deceased}      &    &        &  2 & 0.1\% &    \\ \hline
\multicolumn{1}{l|}{}  & 2046   &        &  2046 & \\
\end{tabular}
\end{table}
For sake of brevity we do not report a table with all the tag changes, but by observing Table~\ref{tab:triagetag}, it is clear that, some color tags are changed to another tag (always an adjacent tag).
\par
In Tables~\ref{tab:distroutcomes} we report the distribution of patients on the basis of the discharge tag and according to the list of the outcomes.
\begin{table}[htbp]
\caption{Distribution of patients on the basis of the discharge tag and according to the actual list of the outcomes.}\label{tab:distroutcomes}
\centering
\begin{tabular}{l|c|c|c|c|c|c}
   & {\sc White} & {\sc Green} & {\sc Yellow} & {\sc Red} &   \\ \hline
  \emph{O1} & 121 & 1025 & 23 &   & 1169 \\
  \emph{O2} & 39 & 496 & 21 &  & 556 \\
  \emph{O3} & & 29 & 182 & 14  &  225 \\
  \emph{O4} & &  & 4  & 4 &   8 \\
  \emph{O5} & & 19 & 10  &   &  29 \\
  \emph{O6} & 3 & 17 & 2  &   &  22 \\
  \emph{O7} & 8 & 26 & 1  &   &  35 \\
  \emph{O8} &  &  &   &   2 & 2 \\
  \emph{O9} &  &  &   &  &  -  \\ \hline
   & 171 & 1612 & 243 & 20  & 2046 \\
\end{tabular}
\end{table}

\section{The Discrete Event Simulation model}\label{sec:desm}
In this section we detail the DES model of the ED reported in Section~\ref{sec:casestudy}. An entity is created on the patient arrival. This event represents the entry of the patient into the system under consideration. Then the entity flows through the different segments of the model according to specified logical rules. These latter enable to reproduce the patient flow described in Section~\ref{sec:casestudy}. At the end of the DTCP, the entity is discharged from the system model.
\par
We implemented the simulation model by means of {\sf ARENA~15.1} Simulation Software \cite{ARENA,simulationwitharena}.
\par
As regards the KPIs of interest, to meet specific demand of the ED managers, we focus on the analysis of the patient flow from the beginning of the DTCP and, in particular, in monitoring, for each color tag,
\begin{itemize}
\item the \textit{Waiting Time} (\textbf{WT}) between the end of the triage and the starting of the visit, namely $t_2-t_1$ in Figure~\ref{fig:times};
\item the \textit{Total Time} (\textbf{TT}) after the triage, i.e. the time between the end of the triage and the discharge, namely $t_5-t_1$ in Figure~\ref{fig:times}.
\end{itemize}
Note that, of course, TT does not coincide with LOS, since the initial part of the pathway until the end of the triage is not considered. This choice is motivated by the request of the practitioners of focusing on all the processes following the triage phase.

\subsection{Input analysis}
\begin{figure}[htbp]
  \centering
    \includegraphics[width=8.5truecm]{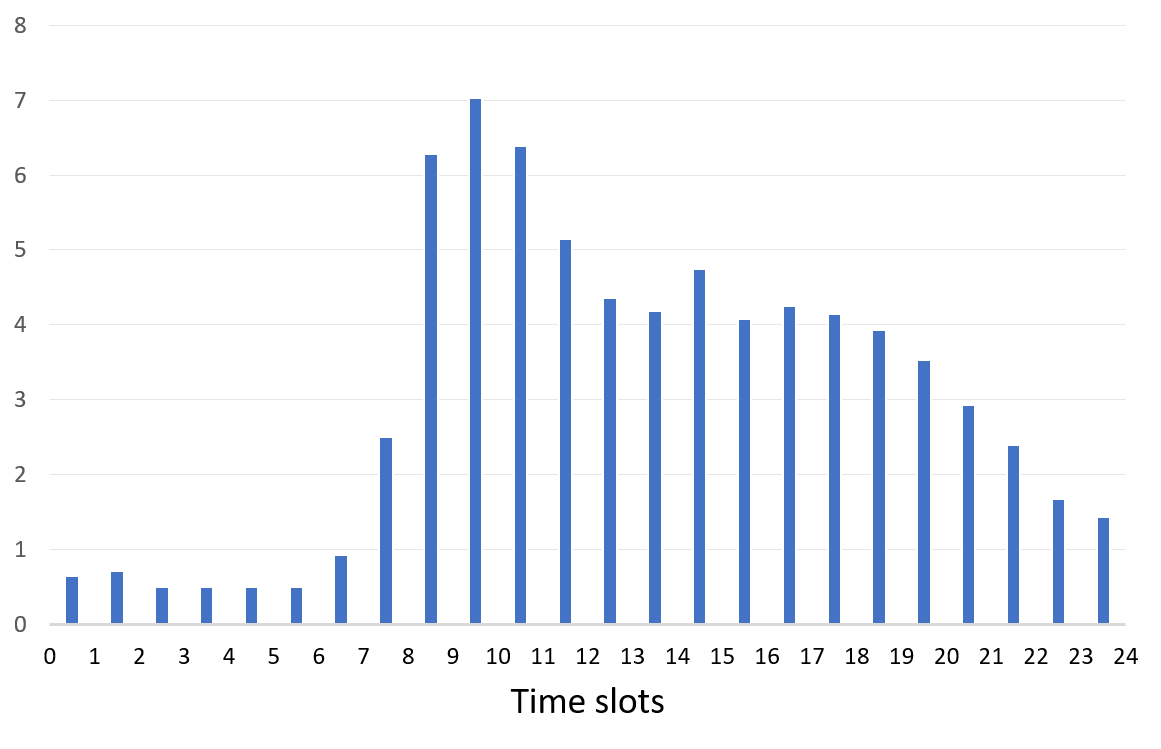}
  \caption{Plot of the hourly arrival rates.}\label{fig:arrivals}
\end{figure}

The data were used for a detailed input analysis of all the processes in the ED. As regards the arrival process, we adopt the standard assumption used in literature that the arrival process to an ED is a Nonhomogeneous Poisson Process (NHPP) (see \cite{Whitt.2017,Kuo.2016,Zeinali.2015,Ahmed.2009,Ahalt.2018,Guo.2017}). In fact, the adoption of a nonhomogeneous process is necessary since patients interarrival times are strongly affected by the arrival hour. In order to obtain a good accuracy of the arrival rate, we consider 24 time slots for each day at an hourly basis, starting from 00:00. Therefore, by using a standard procedure (see e.g. \cite{law:15}), we approximate the arrival rate function by a piecewise constant function. A plot of the hourly arrival rate is reported in Figure~\ref{fig:arrivals}.
Thanks to an ARENA built-in tool (see \cite{simulationwitharena}) it is possible to generate entity arrivals according to
a nonstationary Poisson process.
\par
As regards the timing employed in the \emph{visit} process and in the \emph{additional examination} process, by using the collected data we obtain the probability distribution of the times (in minutes) as reported in Table~\ref{tab:timesvisit} for each color tag.
\begin{table}[htbp]
\caption{Probability distribution of \emph{visit} and \emph{additional examination} times.}\label{tab:timesvisit}
\centering
\begin{tabular}{l|l|l}
            & VISIT & ADDIT. EXAMS \\ \hline
{\sc White} & Lognormal(7.87,9.76) & Exp(44.4)\\
{\sc Green} & Lognormal(11.7,10.6) & Exp(80.4)\\
{\sc Yellow} & Normal(15.2,7.72) &  Weibull(236,0.731)\\
{\sc Red} & $8+$Gamma(10.2,1.49)& Weibull(86.9,0.678) \\ \hline
\end{tabular}
\end{table}
As concerns the resources used, each patient requires one physician, one nurse and one site in the room corresponding to the color tag.

\subsection{Model validation}
In order to guarantee that the DES model we built provide us with a sufficient accuracy of the output, the model has been widely verified and validated. As regards the model validation, we compared the real system values with the corresponding simulation outputs, namely the average values (with their confidence interval) obtained from 50 independent simulation replications each of them one month long, with a warm up period of 24 hours. In particular, we consider some fundamental KPIs of the overall process in terms of \emph{times} and \emph{entity counters}.
\par
\begin{figure}[htbp]
  \centering
    \includegraphics[width=8.5truecm]{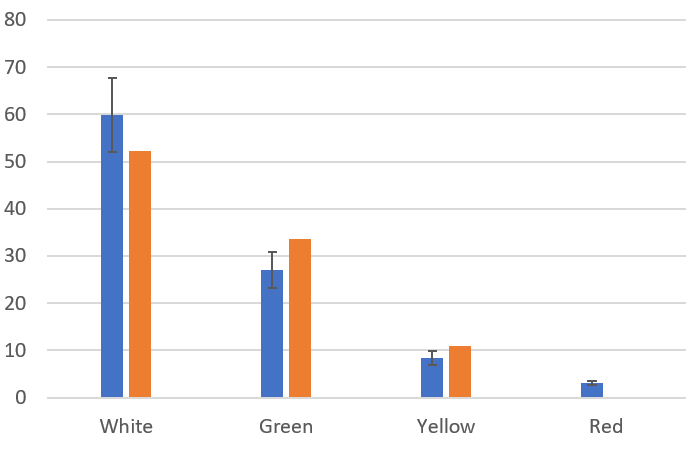}
  \caption{Plot of current values (in orange) and simulation output (in blue) of WT (in minutes) with the confidence interval.}\label{fig:validationwt}
\end{figure}
\begin{figure}[htbp]
  \centering
    \includegraphics[width=8.5truecm]{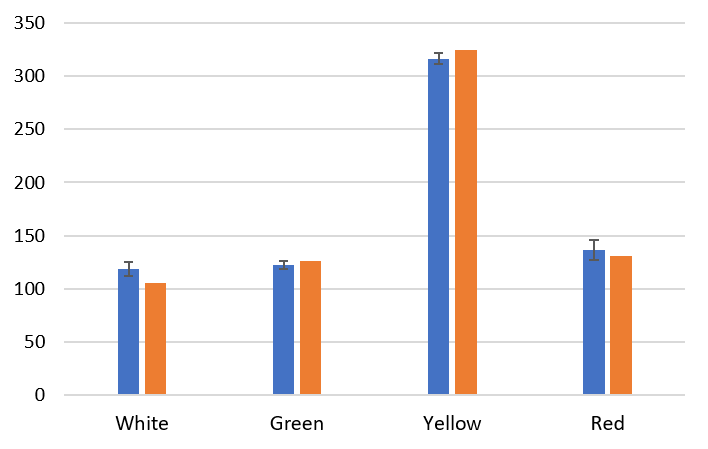}
  \caption{Plot of current values (in orange) and simulation output (in blue) of TT (in minutes) with the confidence interval.}\label{fig:validationtt}
\end{figure}

As concerns times, we focus on the waiting time WT and on the total time TT previously defined. In Figure~\ref{fig:validationwt} and Figure~\ref{fig:validationtt} we report the current values, namely those corresponding to the ``as--is'' status, and the simulation output of WT and TT (in minutes) with the relative confidence interval (with 95\% confidence level).
These plots clearly evidence the reliability and the good accuracy of the simulation model. Indeed, the simulation output represents a good approximation of the current values for all color tags.
\par
As regards the entity counters, we compare the current values of the outcomes as reported in
Table~\ref{tab:distroutcomes}, with the corresponding outputs of the simulation model reported in Table~\ref{tab:distroutcomesvalidation} (for the sake of brevity, we do not report the corresponding plots).
\begin{table*}[htbp]
\caption{Output values (with confidence interval) for the outcomes returned by the simulation.}\label{tab:distroutcomesvalidation}
	\centering
	\begin{tabular}{l|c|c|c|c}
		& {\sc White} & {\sc Green} & {\sc Yellow} & {\sc Red}\\ \hline
		O1 & $120.53\pm2.22$ & $1019.74\pm5.88$ & $23.35\pm0.91$ & \\
		O2 & $39.41\pm1.29$ & $489.85\pm4.69$ & $21.45\pm1.02$ &\\
		O3 & & $28.83\pm1.15$ & $179.18\pm3.08$ & $13.51\pm0.73$\\
		O4 & &  & $3.79\pm0.42$  & $4.18\pm0.41$\\
		O5 & & $18.14\pm0.87$ & $10.24\pm0.65$  &\\
		O6 & $3.15\pm0.32$ & $16.93\pm0.76$ & $2.09\pm0.28$  &\\
		O7 & $7.06\pm0.60$ & $23.63\pm1$ & $1.64\pm0.28$ &\\
		O8 &  &  &   & $2.16\pm0.30$ \\ \hline
	\end{tabular}
\end{table*}
A comparison between the two tables clearly evidences the reliability and the good accuracy of the simulation model. In particular, the model output values corresponding to each outcome and to each color tag are an accurate approximation of the current values, taking into account the confidence interval.
\par
%

\section{Design of experiments and results}\label{sec:results}
In this section, we report experimental results obtained by our DES model. Our aim is to determine performance measures of the ED, considering different scenarios, in order to evaluate possible policy changes leading to an overall improvement. To this aim, we consider hypothetical scenarios where patient arrival rate is artificially changed. In particular, we preliminarily consider an increase of a prefixed percentage of the arrival rate due to the growth in demand and a mildly loaded situation, namely a gradual increase of patient arrivals over a period of few days of the week. Then we turn to the main focus of this work, namely extremely loaded situations, possibly due to some critical conditions, for instance a natural disaster.
For each scenario, we assess how the ED response times change. In particular, we consider the KPIs of interest defined in Section~\ref{sec:desm}, i.e. the waiting time  WT and the total time TT. Moreover, we monitor the resources usage.
More specifically, we monitor the instantaneous utilization (on hourly basis) of the resources. 
Note that instantaneous utilization is preferred to 
average utilization since high utilization can cause saturation and performance deterioration, even though utilization is low when averaged over a long interval. We used VBA (Visual Basic for Applications) ARENA modules for collecting data concerning the instantaneous utilization of the resources so that we have been able to process them.
In particular, we consider the usage of the green and yellow areas.
We recall that both green and white tagged patients are assigned to the green area.

\subsection{Increase of a prefixed percentage of the arrival rate}\label{subsec:prefixed}
\begin{figure}[htbp]
  \centering
    \includegraphics[width=8.5truecm]{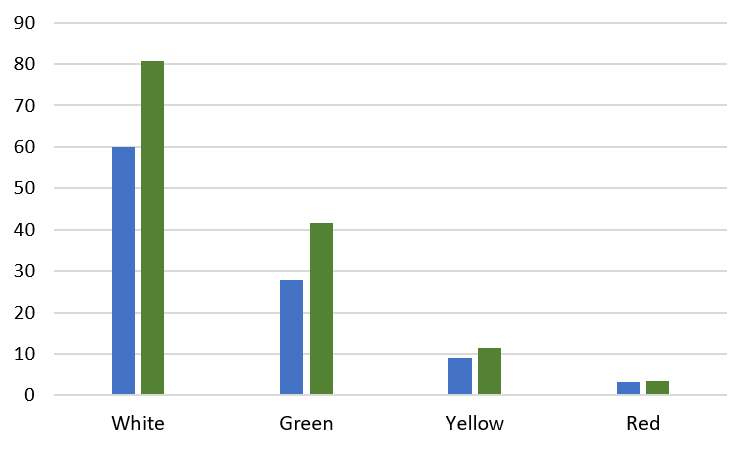}
  \caption{WT (in minutes): plot of the comparison between the current ``as--is'' status (in blue) and a prefixed percentage increase of the arrival rate (in green).}\label{fig:scenario1a}
\end{figure}

\begin{figure}[htbp]
  \centering
    \includegraphics[width=8.5truecm]{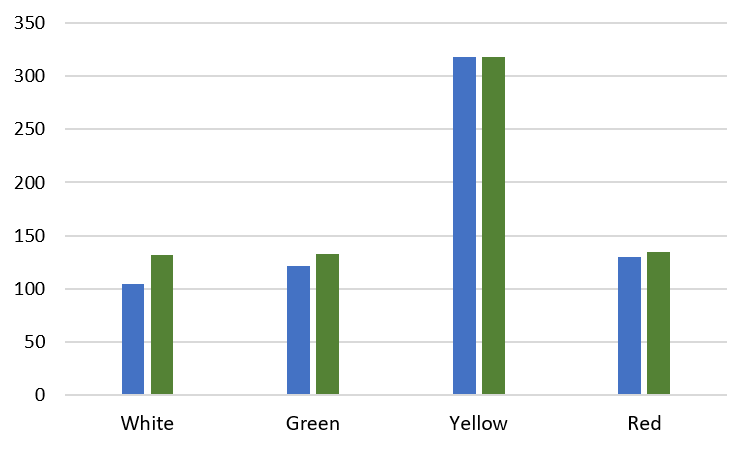}
  \caption{TT (in minutes): plot of the comparison between the current ``as--is'' status (in blue) and a prefixed percentage increase of the arrival rate (in green).}\label{fig:scenario1b}
\end{figure}

\begin{figure}[htbp]
  \centering
    \includegraphics[width=8.5truecm]{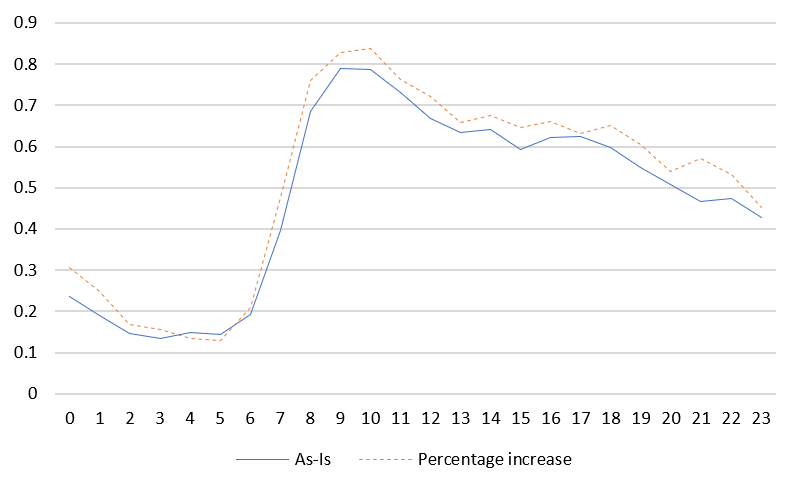}
  \caption{Plot of the \textit{usage of the green area}.}\label{fig:scenario1c}
\end{figure}

\begin{figure}[htbp]
  \centering
    \includegraphics[width=8.5truecm]{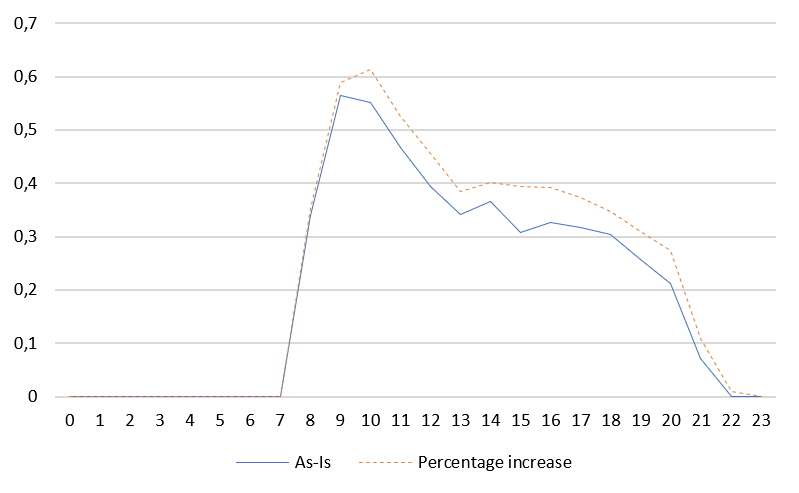}
  \caption{Plot of the \textit{usage of the yellow area}.}\label{fig:scenario1d}
\end{figure}

We consider an increase of $10\%$ of the hourly arrival rate with respect to the current value. We run 10 independent replications and the length of each replication is one month with a warm up period of 24 hours. In Figures~\ref{fig:scenario1a}-\ref{fig:scenario1b} we report the comparison in terms of WT and TT, respectively.
Of course, the uniform increase of the demand implies longer waiting/stay times. However, this grow of the arrival rate does not significantly affect waiting/stay times for yellow and red tagged patients. This is mainly due to the priority criterion and also to the small number of red and yellow tagged patients arriving with respect to green and white ones. 
In Figure~\ref{fig:scenario1c}-\ref{fig:scenario1d} we report the comparison between the usage of the green and the yellow areas over the 24 hours, namely the instantaneous utilization (on hourly basis) of each of them. We do not consider the usage of the shock room because of the reduced number of red tagged patients.
From Figure~\ref{fig:scenario1c} it can be observed that the usage of the green area in the rush hours is close to one even in the ``as--is'' status and that, as expected, the percentage increase causes a corresponding uniform growing of the utilization. Before discussing the usage of the yellow area, note that since only one physician is on duty during the night, actually only the green area is used in the night for treatment of both green and yellow tagged patients. Therefore, as shown in  Figure~\ref{fig:scenario1d}, no patient is visited in the yellow area during the night
while throughout the day, when two physicians are on duty, the area is strongly used and a uniform increase of the usage is observed, accordingly to the grow of the average hourly arrival rate.

\subsection{A mildly and an extremely loaded scenario}
Now we consider unexpected conditions, namely a spike of patient arrival rate due to a sudden and critical event (for instance, in the extreme case, an earthquake). We focus on two possible artificial scenarios: a mildly and an extremely loaded scenario corresponding to two different unpredictable occurrences. For each experiment, we run 10 independent replications and the length of each replication is 5 days, with a warm up period of 24 hours. The choice of using a 5 days length is motivated by the need of accurately monitor the actual effect of arrival spikes. More precisely, the aim is to avoid that, in the computation of the average values for those KPIs which are computed as an average over the whole replication length, too many occurrences concerning standard days (without spikes) are included. Indeed, this could cause a not negligible bias in the results. 
\par
As regards the mildly loaded situation, we adopt a gradual increase/decrease of the arrival rate over three days of the week (starting from day 2). More precisely, similarly to \cite{Ahalt.2018}, we increase the arrival rate from 5\% to 25\%, depending on time slots, according to the scheme in Table~\ref{tab:incrementigraduali}.
\begin{table}[htbp]
\caption{Percentage increases of the arrival rate for \emph{Day~2}, \emph{Day~3} and \emph{Day~4} for different time slots}\label{tab:incrementigraduali}
	\centering
	\begin{tabular}{r|c|c|c}
 & \emph{Day 2} & \emph{Day 3} & \emph{Day 4}  \\\hline
00:00 -- 08:00 & $+5\%$ & $+20\%$ & $+20\%$ \\
08:00 -- 14:00 & $+10\%$ & $+25\%$ & $+15\%$ \\
14:00 -- 20:00 & $+15\%$ & $+25\%$ & $+10\%$ \\
20:00 -- 24:00 & $+20\%$ & $+20\%$ & $+5\%$ \\ \hline
	\end{tabular}
\end{table}
\par
\begin{figure}[htbp]
  \centering
    \includegraphics[width=8.5truecm]{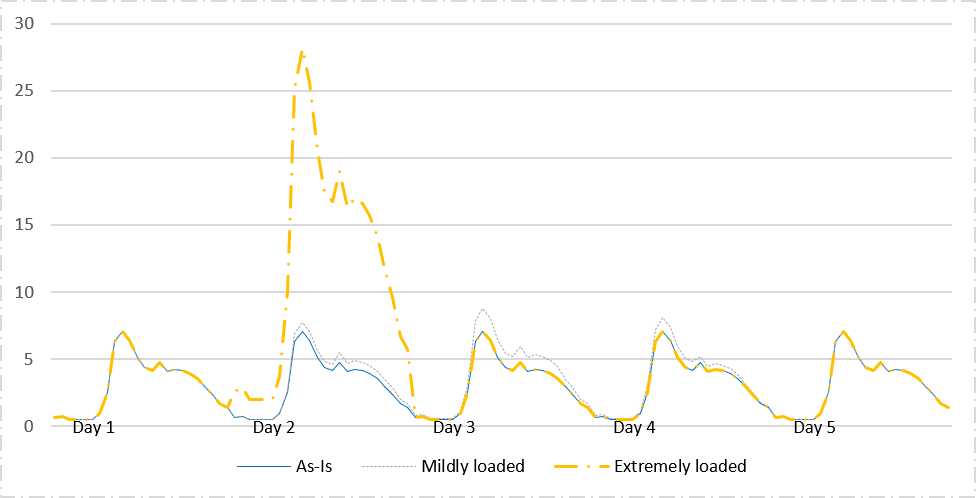}
  \caption{Plot of the increased patients arrival rate (midly loaded in grey, extremely loaded in yellow) and the unmodified arrival rate (in blue). }\label{fig:arriviaumentati}
\end{figure}

As concerns the extremely loaded scenario, we try to reproduce a major emergency. To this aim, we consider a 300\% increase of the arrival rate centered over the 24 hours of the Day~2 of the week. Figure~\ref{fig:arriviaumentati} reports the increased hourly arrival rate for both scenarios along with the unmodified arrival rate.
\par
\begin{figure}[htbp]
  \centering
    \includegraphics[width=8.5truecm]{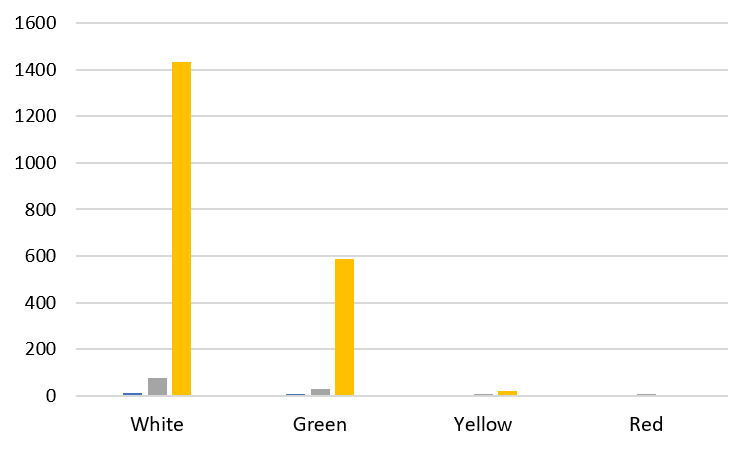}
    \includegraphics[width=8.5truecm]{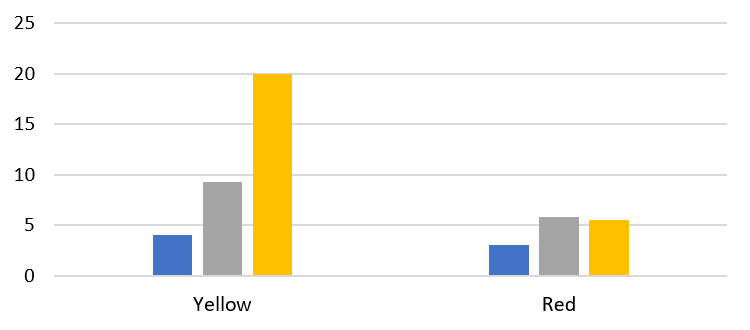}
  \caption{WT (in minutes): plot of the comparison between the current ``as--is'' status (in blue) and the mildly (in grey) and extremely (in yellow) loaded scenarios. In the bottom the detail for yellow and red tagged patients. }\label{fig:scenario2a}
\end{figure}
\begin{figure}[htbp]
  \centering
    \includegraphics[width=8.5truecm]{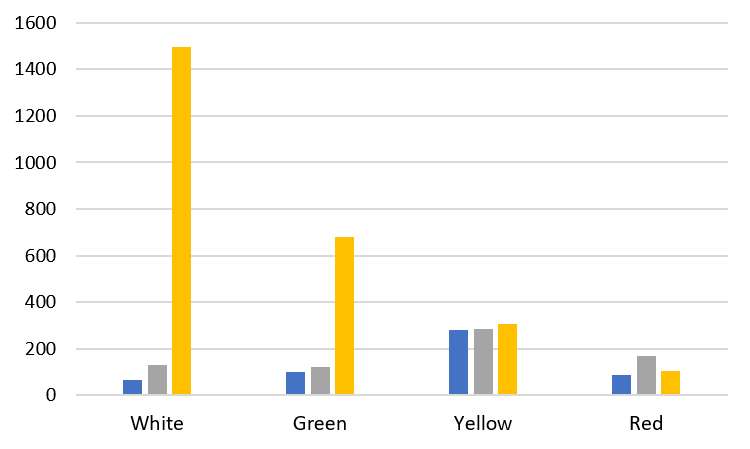}
  \caption{TT (in minutes): plot of the comparison between the current ``as--is'' status (in blue) and the mildly (in grey) and extremely (in yellow) loaded scenarios.}\label{fig:scenario2b}
\end{figure}

In Figure~\ref{fig:scenario2a} we report the comparison between the current ``as--is'' status and the two scenarios in terms of WT.
Similarly, in Figure~\ref{fig:scenario2b} the same comparison is reported in terms of TT.
As expected, we observe a huge increase for both WT and TT in the extremely loaded scenario, for low--complexity patients (white and green tagged): the WT would exceed one day for white tagged patients and 10 hours for the green tagged ones and this is not acceptable. As regards the mildly loaded scenario, a moderate increase is highlighted, showing that both WT and TT are actually still feasible. A different outcome is pointed out for higher complexity patients. As regards the red tagged ones, in both the scenarios we do not change their current percentage, with respect to the other color tagged ones, given in Table~\ref{tab:triagetag}
(scenarios with changes in this percentage are reported afterwards). In this case, even a huge increase of the overall arrivals does not lead to exceed one red tagged patient arrival per hour. Therefore both WT and TT does not grow significantly, also due to the high priority assigned to these patients. Similar result is observed for the yellow tagged patients. Moreover, note that the WT for red tagged patients is still approximately zero, in accordance to their high urgency level.
\par
\begin{figure}[htbp]
  \centering
    \includegraphics[width=8.5truecm,height=4.5truecm]{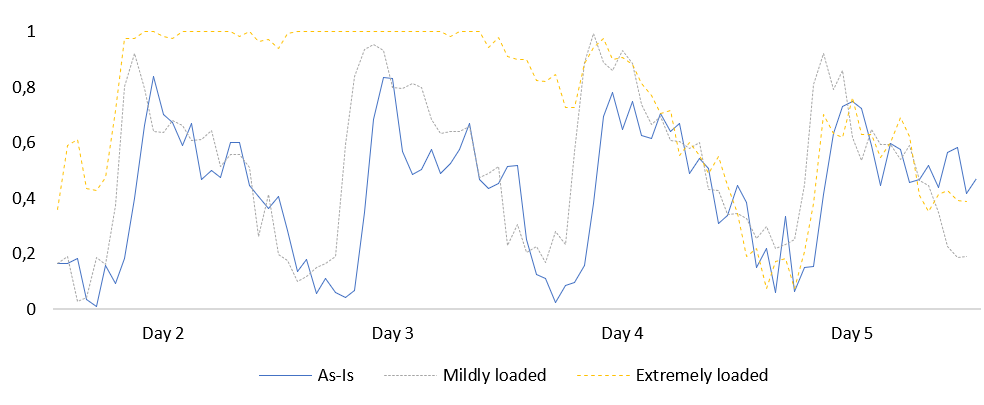}
  \caption{Plot of the \textit{usage of the green area}: the current ``as--is'' status (in blue), the mildly loaded (in grey) and the extremely loaded (in yellow) scenarios.}\label{fig:scenario2c}
\end{figure}
\begin{figure}[htbp]
  \centering
    \includegraphics[width=8.5truecm,height=4.5truecm]{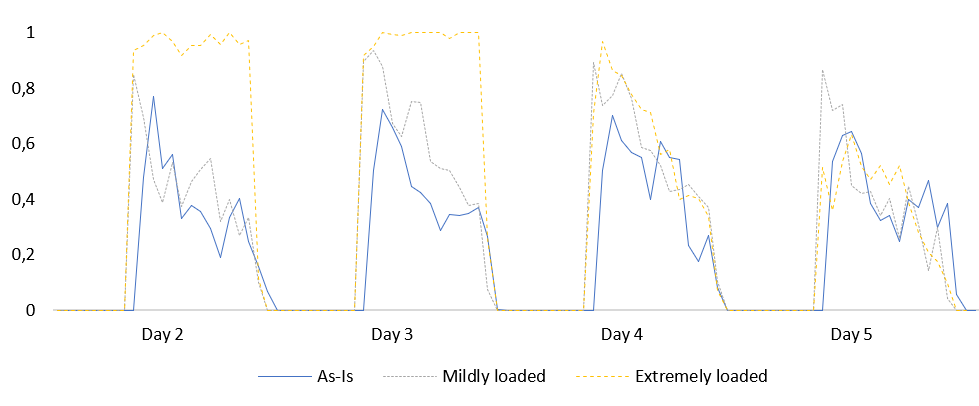}
  \caption{Plot of the \textit{usage of the yellow area}: the current ``as--is'' status (in blue), the mildly loaded (in grey) and the extremely loaded (in yellow) scenarios.}\label{fig:scenario2d}
\end{figure}

Figures~\ref{fig:scenario2c}-\ref{fig:scenario2d} report the comparison between the current ``as--is'' status and the two artificial scenarios in terms of usage of green and yellow areas.
From both these figures, it can be observed how, in the two scenarios, the peak patient arrivals causes a sudden increase of the usage of both the resources (green and yellow areas). In the case of extremely loaded scenario, the peak causes resources saturation even immediately before and after the peak center. 
Note how this phenomenon can be observed only by monitoring the instantaneous resource utilization rather than the average utilization.
\par
Now we analyze more in detail the extremely loaded scenario. Indeed, due to the unpredictability of the phenomenon of peak arrivals for to critical events, it is difficult to create artificial scenarios that actually reproduce what could happen in reality. Therefore, we now analyze some variants of the $300\%$ increase of the arrival rate already considered. In particular, we report the comparison of the current ``as--is'' status with respect to increases of $100\%$, $200\%$, $400\%$ of the arrival rate (always centered in the Day~2 and with the same percentage distribution of the color tags). In the following Figures~\ref{fig:waitingTimepeakArrivals}--\ref{fig:totalTimepeakArrivals}, the corresponding WT and the TT are reported, along with those obtained for the $300\%$ increase scenario.
\begin{figure}[htbp]
	\centering
    \includegraphics[width=8.5truecm,height=4.5truecm]{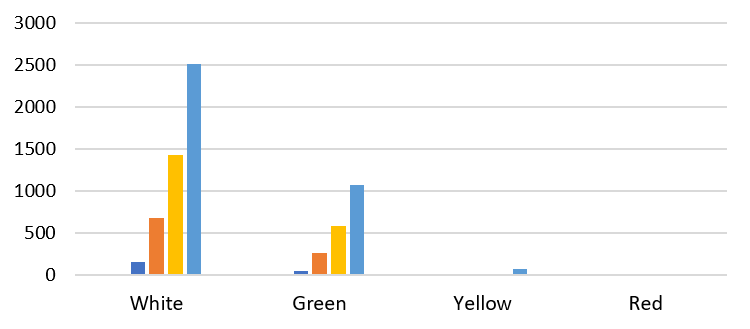}
    \includegraphics[width=8.5truecm]{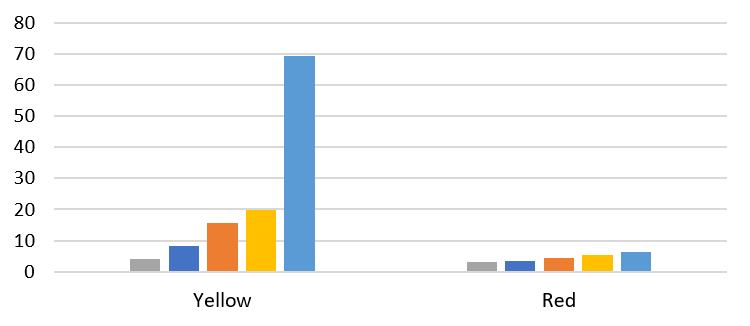}
	\caption{WT (in minutes): plot of the comparison between the current ``as--is'' status (in grey) and the extremely loaded scenario with increases of the arrival rate of $100\%$ (in blue), $200\%$ (in orange), $300\%$ (in yellow) and $400\%$ (in light blue). In the bottom the detail for yellow and red tagged patients.}
	\label{fig:waitingTimepeakArrivals}
\end{figure}
\begin{figure}[htbp]
	\centering
	\includegraphics[width=8.5truecm,height=4.5truecm]{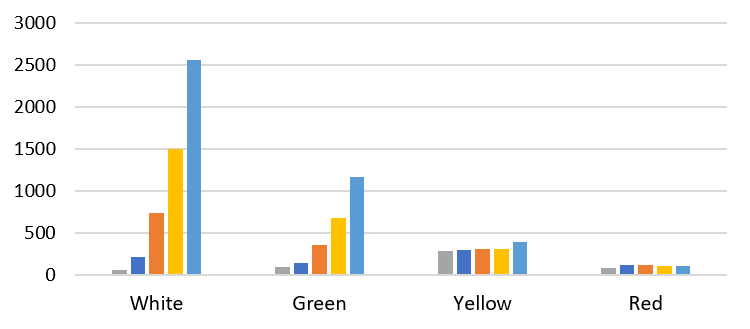}
	\caption{TT (in minutes): plot of the comparison between the current ``as--is'' status (in grey) and the extremely loaded scenario with increases of the arrival rate of $100\%$ (in blue), $200\%$ (in orange), $300\%$ (in yellow) and $400\%$ (in light blue)}\label{fig:totalTimepeakArrivals}
\end{figure}
From Figure~\ref{fig:waitingTimepeakArrivals},
it can be observed, how WT for the red tagged patients still remain acceptable for all the four scenarios, and this is due to the low percentage of red tagged patients arrivals. As regards the yellow tagged ones, WT remains below 10 minutes only for the $100\%$ increase. As concerns the white and green tagged patients, WT becomes unacceptable even with the $100\%$ increase. The TT reported in Figure~\ref{fig:totalTimepeakArrivals} are direct consequence of the corresponding WT.
\par
\begin{figure}[htbp]
	\centering
	\includegraphics[width=8.5truecm,height=5truecm]{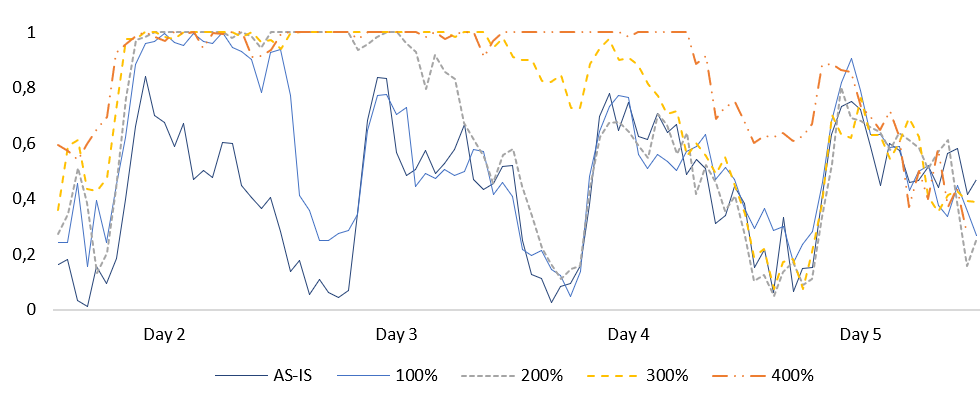}
	\caption{Plot of the \textit{usage of the green area}: the current ``as--is'' status and the extremely loaded scenarios.}
	\label{fig:usageGreenpeakArrivals}
\end{figure}
\begin{figure}[htbp]
	\centering
	\includegraphics[width=8.5truecm,height=5truecm]{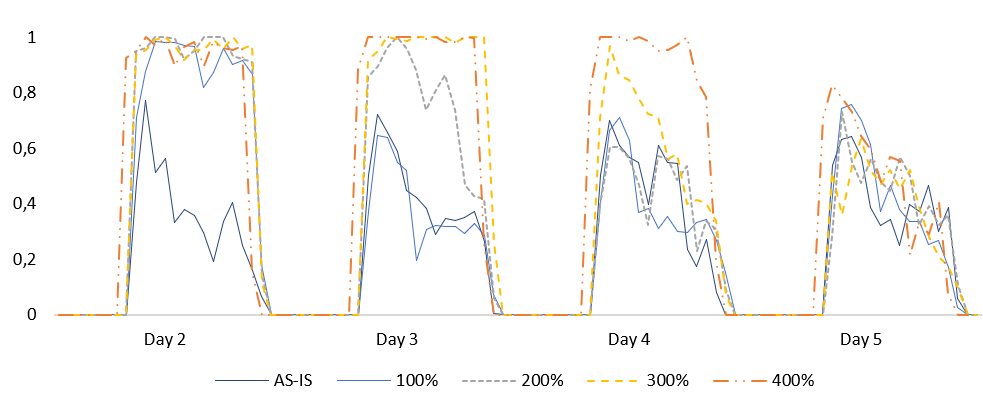}
	\caption{Plot of the \textit{usage of the yellow area}: the current ``as--is'' status and the extremely loaded scenarios.} \label{fig:usageYellowpeakArrivals}
\end{figure}

The same comparison between the current ``as--is'' status and the increases of $100\%$, $200\%$, $400\%$ of the arrival rate is reported in the Figure~\ref{fig:usageGreenpeakArrivals}--\ref{fig:usageYellowpeakArrivals} in terms of resources usage for the green and the yellow areas, respectively.
These figures clearly highlight how, as expected, the percentage increase of the patient arrival rate strongly affect the utilization of both the green and the yellow areas. Note that the usage of the green area also depends on the fact that during the night it is also used for yellow tagged patients since the yellow area is not in use during the night. Both the resources reach saturation points even in the most cautious scenario ($100\%$ increase).
\par\medskip
All the scenarios up to now analyzed are based on increases of the patient arrival rate, keeping unchanged the percentage distribution of different colors tagged patients. Actually, during a critical event, it is also likely to assume that the percentage of high priority patients grows during the peak arrivals. During the latest earthquake (August 24, 2016) that hit Central Italian regions (where the ED considered in this work is located), during the hour corresponding to the maximum peak arrival, up to 14 patients with trauma due to crushing (i.e. red tagged patients) requested assistance. As already mentioned in Section~\ref{sec:gen}, in these cases, i.e. in case of the so called ``maxi--emergency'',  according to current regulations, Italian EDs adopt the ``Internal Emergency Plan for Massive Inflow of Injured'' (we use the Italian acronym PEIMAF). This implies further resources availability and different operating rules aimed at providing an adequate and timely assistance to all the patients who require it. Of course, such a plan cannot be tested during the normal ED activity. However, it is really important to check it accurately in order to ensure its effectiveness whenever it should be activated. A natural way for performing such testing is the use of a DES model. Therefore, we used our DES model also to provide some insights, useful for the decision makers, concerning the definition of the PEIMAF.
\par
In the sequel, we report some analyses we performed, aiming at reproducing a critical situation corresponding to the extremely loaded scenario with an increase of $400\%$ of the overall arrival rate and, in addition, an increase of the red tagged patients arrival during the peak. In particular, we assume an increase of the percentage of red tagged patient arrivals, in order to obtain about 14 of these patients arriving in the peak hour, as really occurred during the recent earthquake. Moreover, we assume the adoption of a possible maxi--emergency PEIMAF plan  and compare the KPIs obtained. More specifically, we adopt the following assumptions:
\begin{description}
\item[{\bf A1}] {\em Patients arrivals:}
\begin{itemize}
\item the percentage of color tags assigned at the triage station is modified during the whole peak day, by assuming that $50\%$ of the arrivals are red tagged patients. 
\end{itemize}
\item[{\bf A2}] {\em Maxi--emergency plan (PEIMAF):} 
\begin{itemize}
\item the number of physicians and nurses on duty is doubled during the peak day, starting from the peak hour (10:00 a.m.), then the shifts return to the normal scheme; 
\item green and yellow tagged patients are not admitted to the ED but send to outpatients facilities; 
\item also green and the yellow areas can be used for treatment of red tagged patients.
\end{itemize}  
\end{description}
In this manner, by {\bf A1} we obtain about 14 red tagged patient arrivals in the peak hour trying to reproduce a really occurred critical situation. By {\bf A2} we implement a possible simple configuration of a maxi--emergency plan, only for experimental purpose. Actually, the operational procedures provided by a PEIMAF plan are much more complex and articulate and here we only report an illustrative example to show how
our DES model can be fruitfully used to test a maxi--emergency plan.
\par
\begin{figure}[htbp]
	\centering
	\includegraphics[width=8.5truecm,height=5truecm]{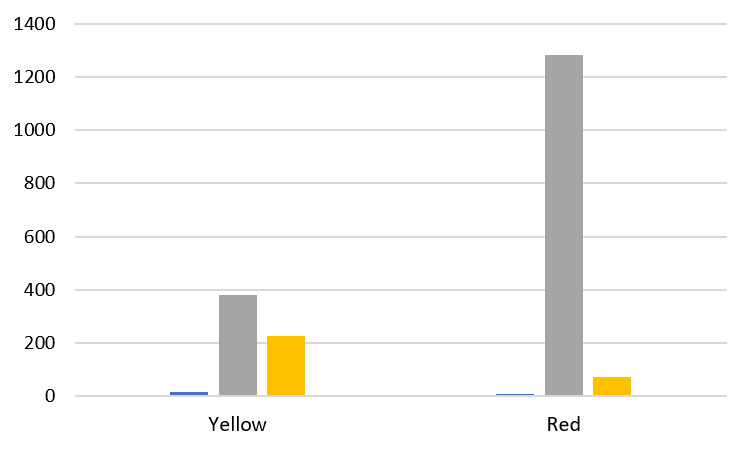}
	\caption{WT (in minutes): plot of the comparison between the current ``as--is'' status (in blue) and the extremely loaded scenario with the modified percentage of red tags assigned at the triage as in {\bf A1}, with (in yellow) and without (in grey) adopting a maxi--emegency plan as in {\bf A2}.}
	\label{fig:PEIMAF-waitingTime}
\end{figure}
\begin{figure}[htbp]
	\centering
	\includegraphics[width=8.5truecm,height=5truecm]{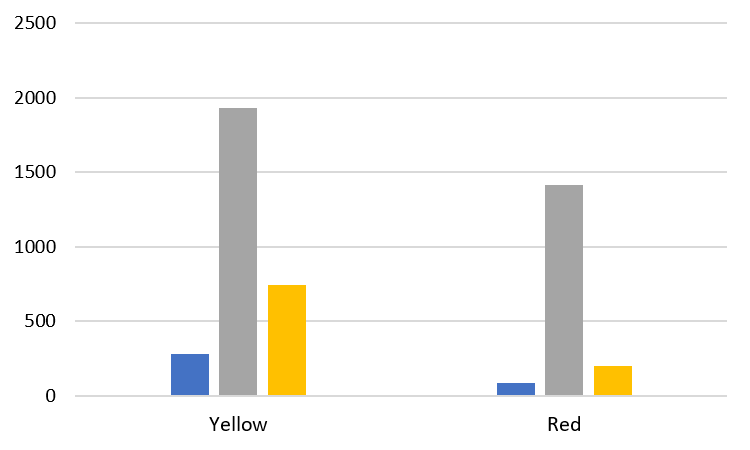}
	\caption{TT (in minutes):plot of the comparison between the current ``as--is'' status (in blue) and the extremely loaded scenario with the modified percentage of red tags assigned at the triage as in {\bf A1}, with (in yellow) and without (in grey) adopting a maxi--emergency plan as in {\bf A2}.}
	\label{fig:PEIMAF-totalTime}
\end{figure}
\begin{figure}[htbp]
	\centering
	\includegraphics[width=8.5truecm,height=5truecm]{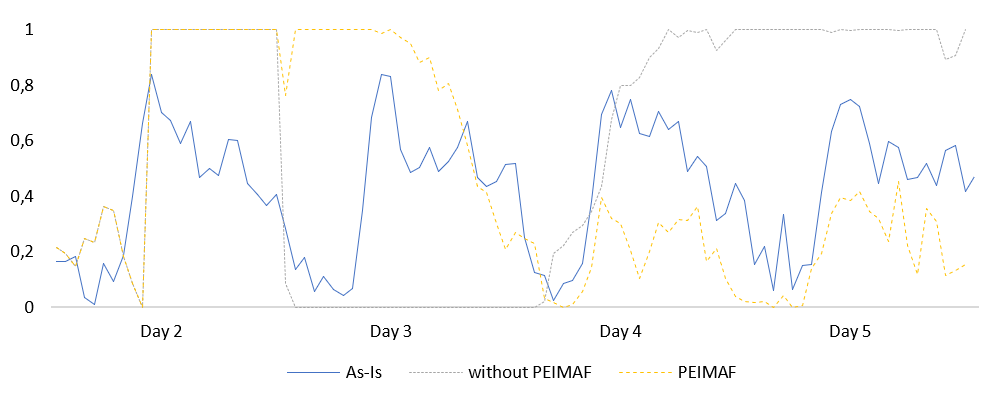}
	\caption{Plot of the \textit{usage of the green area}: the current ``as--is'' status and the extremely loaded scenario with and without the maxi--emergency plan.}
	\label{fig:PEIMAF-usageGreen}
\end{figure}
\begin{figure}[htbp]
	\centering
	\includegraphics[width=8.5truecm,height=5truecm]{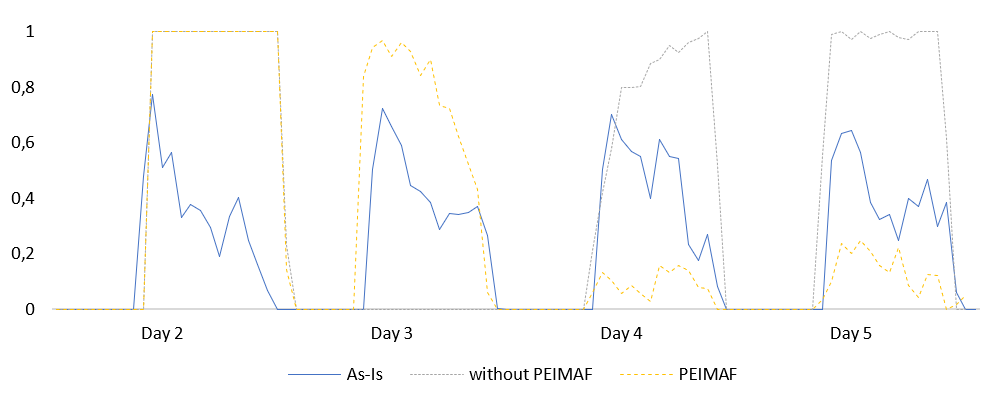}
	\caption{Plot of the \textit{usage of the yellow area}: the current ``as--is'' status and the extremely loaded scenario with and without the maxi--emergency plan.} \label{fig:PEIMAF-usageYellow}
\end{figure}
\begin{figure}[htbp]
	\centering
	\includegraphics[width=8.5truecm,height=5truecm]{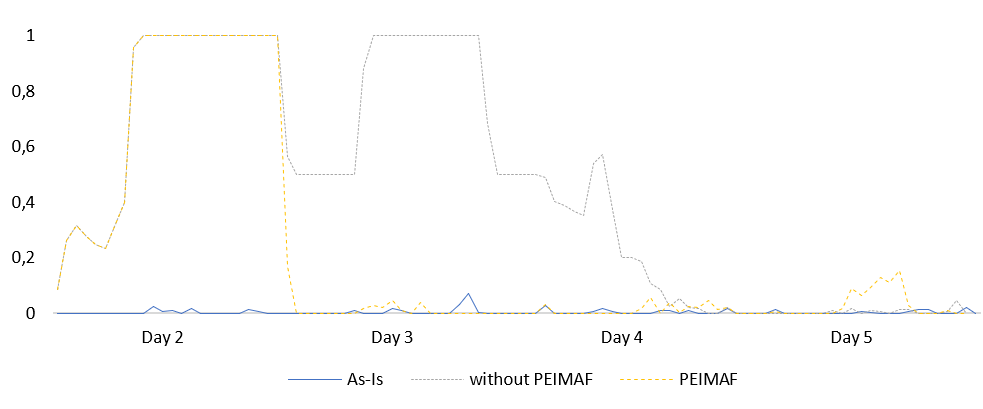}
	\caption{Plot of the \textit{usage of the shock room}: the current 
	``as--is'' status and the extremely loaded scenarios with and without the maxi--emergency plan.} \label{fig:PEIMAF-usageRed}
\end{figure}

Figures~\ref{fig:PEIMAF-waitingTime}--\ref{fig:PEIMAF-totalTime} report the WT and the TT for the ``as--is'' status and the extremely loaded scenario with $400\%$ increase of patient arrival rate and the percentage of red tags assigned at the triage station modified according to {\bf A1}. The comparison concerns the values of these KPIs obtained without adopting a maxi--emergency plan and by using the plan described in {\bf A2}.
The figures clearly evidence the huge times resulting without adopting the maxi--emergency plan. Even by adopting the plan as specified above in {\bf A2}, WT  exceeds 3 hours for yellow tagged patients and 1 hour for the red tagged ones and, obviously, this is unacceptable. This situation is confirmed by the plot of the usage of green and yellow areas and shock room reported in Figures~\ref{fig:PEIMAF-usageGreen}--\ref{fig:PEIMAF-usageRed}.
First, note that in case of emergency plan not activated, and hence only two physicians are on duty during the peak day, they both are assigned to the shock room and therefore, the green and the yellow areas cannot be used. On the contrary, if the emergency plan is activated and hence four  physicians are on duty during the peak day, also green and yellow areas can be used even for treatment of higher complexity patients. In any case, even if the adoption of the maxi--emergency plan as assumed in {\bf A2} leads to an improvement, its overall inadequacy is very evident. The resource saturation is reached for long periods, implying that the resources would have extra requests which cannot be satisfied and hence they are queued. Of course, in case the resources (both human and physical ones) cannot be 
further enlarged, patient diversion policy towards neighboring EDs should be adopted. 
\par
To be aware in advance of the real operational capacity of an ED in case of peak arrivals could be really important for the ED managers. In fact, this allows them to define or adjust a maxi--emergency plan, taking into account the specific environmental context, too.
Thanks to the high flexibility of our ARENA implementation of the DES model of the ED under study, an extensive scenario analysis can be performed aimed at assessing the ED operational capacity, namely all the KPIs of interest in many and different real critical situations.

\section{Conclusions}\label{sec:conc}
In this paper we propose a DES model for studying 
the ED of a medium dimension hospital in a region of Center Italy recently hit by a severe earthquake. The aim is to assess how fundamental KPIs change in response to different increase patterns of the patient arrival rate. In particular, we focus on
extremely loaded situations for the ED, due to critical events like a natural disaster. To evaluate the performance of the EDs, time--related measurements have been considered as well as resources usage. 
Several scenarios have been considered, including someones artificially created, trying to reproduce real mass casualty occurrences. The experimental results showed that, when the increase of the arrival rate is low or moderate, the ED performance does not significantly deteriorate.
Instead, in case of extreme events with high patient peak arrivals, the adoption of an exceptional emergency plan is necessary to ensure effective and timely assistance. 
\par
The model proposed in this paper refers to a specific ED, but thanks to the flexibility of its implementation, it can be easily adapted to reproduce the patient flow of other EDs.
We believe that the model we proposed has a twofold merit: on one side it represents an effective decision support system, allowing to assess the performance of the ED under study, highlighting possible operational improvements and enabling decision makers to better allocate ED resources. On the other side, the model can be used to perform scenario analyses to help managers to define in advance maxi--emergency plans which, of course, cannot be experiment during the normal activity of the ED.

%

\begin{acknowledgements}
The authors wish to express their gratitude to Dr. Stefania Mancinelli and to Dr. Marco Pierandrei
from ``Direzione Medica Ospedale di Fabriano'' who enable us to carry out this study. Moreover we thank Dr. Massimo Maurici and Ing. Luca Paulon from ``Dipartimento di Biomedicina e Prevenzione'' and ``Laboratorio di Simulazione e Ottimizzazione dei servizi del SSN'' of the Universit\`a di Roma ``Tor Vergata'', for useful discussions on an early stage of this work.
\end{acknowledgements}

%
\section*{Conflict of interest}
The authors declare that they have no conflict of interest.


\end{document}